\documentclass[a4paper]{jpconf}
\usepackage{graphicx}

\begin{document}
\title{Low-degree modes}

\author{Rafael A. Garc\'\i a}

\address{Laboratoire AIM, CEA/DSM-CNRS - U. Paris Diderot - DAPNIA/SAp, 
91191 Gif-sur-Yvette Cedex, France}

\ead{rafael.garcia@cea.fr}

\begin{abstract}
We review some of the works done during the last year and some of the challenges we have to face today in the study of low-degree acoustic and gravity modes and their implications in the study of the solar internal structure and dynamics.   
\end{abstract}

\section{Introduction}

The quest of the knowledge of the structure and dynamics of the solar interior has been possible thanks to the study of the resonant acoustic (p) modes that are trapped in the solar interior. 

Since the solar rotation lifts the azimuthal degeneracy of the resonant modes, their
eigenfrequencies, $\nu_{n \ell m}$, are split into their $m$-components; where
$\ell$ is the angular degree, $n$ the radial order, and, $m$ the azimuthal
order. This separation ---usually called rotational splitting (or just splitting)--- depends on the rotation rate in the region
sampled by the mode. In the same way, the precise frequency of a mode depends on the physical properties of the cavity where the mode propagates. Using inversion techniques the rotation rate, the sound speed or the density profile at different locations inside the Sun can be inferred from a suitable lineal combination of the measured modes. But, during the last year, a particular effort has been done in the extraction of physical information directly from the combination of frequencies: the large and the small frequency separations. 

\section{Frequency separations}
\begin{figure}
\begin{center}
\includegraphics[width=\hsize]{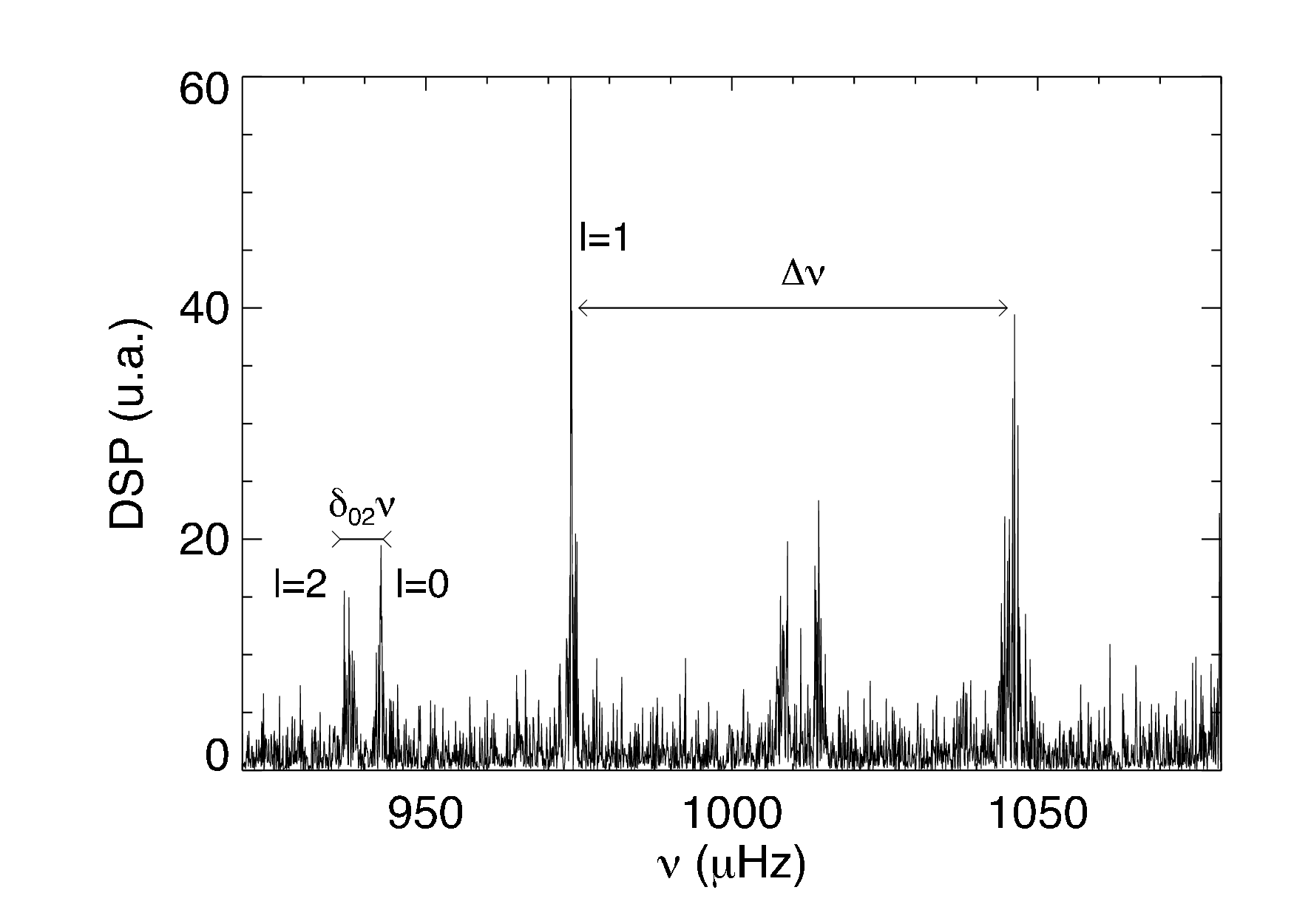}
\end{center}
\caption{\label{Separation} Region of a power spectrum density of a Sun-as-a--star instrument where the large and small frequency separations are shown. }
\end{figure}

The large frequency separation of low-degree p modes is given by (see also Fig.~\ref{Separation}):
\begin{equation}
\Delta \nu_{\ell} (n)= \nu_{n,\ell}-\nu_{n-1, \ell} \;\;\; .
\end{equation}

This large frequency separation gives an idea of the mean quantities of the star (e.g. its mean density) as it depends inversely on the sound-travel time between the center and the surface (see e.g. \cite{JCD2002}):
\begin{equation}
\Delta \nu_{\ell} (n)=\left[2 \int_0^R \frac{dr}{c}\right]^{-1}  \;\;\;,
\end{equation}
where R is the solar radius and c is the sound speed. Recently, Kholikov (these proceedings) using low-degree modes ($\ell \le$ 3) from GONG and MDI data sets has been able to recompute the solar acoustic radius by measuring this large separation. 

The small frequency separation of  low-degree p modes is given by (see also Fig.~\ref{Separation}):

\begin{equation}
\delta \nu_{\ell, \ell+2} (n)= \nu_{n,\ell}-\nu_{n-1, \ell+2} \;\;\; .
\end{equation}

This difference is mainly dominated by the sound-speed gradient in the core and, therefore, it is sensitive to the chemical composition in the central regions and can be used to infer the central hydrogen content (the age of the star). Using the asymptotic theory it can be shown that \cite{1991soia.book..401C}:

\begin{equation}
\delta \nu_{\ell, \ell+2} (n) \simeq -(4\ell + 6) \frac{\Delta \nu_\ell (n)}{4 \pi^2 \nu_{n,\ell}}\int_0^R \frac{dc}{dr} \frac{dr}{r} \;\;\;.
\end{equation}

As the frequencies of both modes are very close, they have similar near-surface effects and the small separation is mostly unaffected by such effects. But some residuals can still remain. This is the reason why the so-called frequency separation ratios (see e.g. \cite{2003A&A...411..215R}) are more useful.

Using the small separation and the frequency separation ratios and by comparing them between different solar models it has been found that models constructed with low metallicity are incompatible with the observations \cite{2007ApJ...655..660B}. These results provided strong support for lowering the theoretical uncertainties on the neutrino fluxes that were recently raised due to the controversy over solar abundances. Using the same 
separation ratios and comparing them with a large Monte Carlo simulation of standard solar models, the mean molecular weight of the solar core and the metallicity of the solar convection zone have been better constrained \cite{2007arXiv0705.3154C}.

\section{Data analyses}
More than 4000 days have been cumulated in the modern helioseismic data sets from ground-based instruments (i.e. BISON\footnote{Birmingham Solar Oscillation network \cite{1996SoPh..168....1C}}, GONG\footnote{Global Oscillation Network Group \cite{HarHil1996}}), as well as those placed aboard the space based SoHO\footnote{Solar and Heliospheric Observatory \cite{DomFle1995}} mission (GOLF\footnote{Global Oscillations at Low Frequency \cite{GabGre1995}}, VIRGO\footnote{Variability of solar IRradiance and Gravity Oscillations\cite{1995SoPh..162..101F}} and SOI/MDI\footnote{Solar Oscillations Investigation/Michelson Doppler Imager\cite{1995SoPh..162..129S}}). 

Thanks to the extremely high quality of these continuous observations we have achieved a better frequency resolution while the reduction of the background level opened the possibility to measure new low-degree low-order p modes (see for example Salabert et al. these proceedings). Moreover the combination of contemporaneous data sets from different instruments and the application of joint probability has been studied and several modes have been found down to $\sim$ 1 mHz \cite{2007MNRAS.379....2B}.

At the same time, the better quality of the data introduces higher requirements in the precision of the peak-bagging codes and a higher sense on the reliability of the results. Thus, new coordinated efforts have been developed in the community, like the solarFLAG\footnote{solarFLAG URL http://bison.ph.bham.ac.uk/ ~wjc/Research/FLAG.html} collaboration, in order to check the different peak-bagging codes and to study the nature of the possible biases in the extraction of the p-mode parameters. Indeed, it has been shown that the peak  asymmetry of the low-degree modes can be properly extracted in the case of the modes $\ell$=1 and 3 but it could be biased for the pair of modes $\ell$=0 and 2 (Chaplin et al. these proceedings).

More than 11 years of data also means the possibility of studying a full solar activity cycle from both observables, velocity and intensity, opening new challenges. For example, it has been found that the peak asymmetry of the modes changes with the solar cycle only in velocity \cite{JimCha2007}.

\section{Extracting the rotation in the radiative zone: improvements in the splitting measurements}

To extract more information on the rotation in the deepest solar layers we need, on one hand, a better determination of the p-mode splittings because they penetrate deeper inside the core of the Sun. On the other hand, we also need  to detect mixed and gravity modes that mainly propagates inside the radiative zone. 

\begin{figure}
\begin{center}
\includegraphics[width=\hsize]{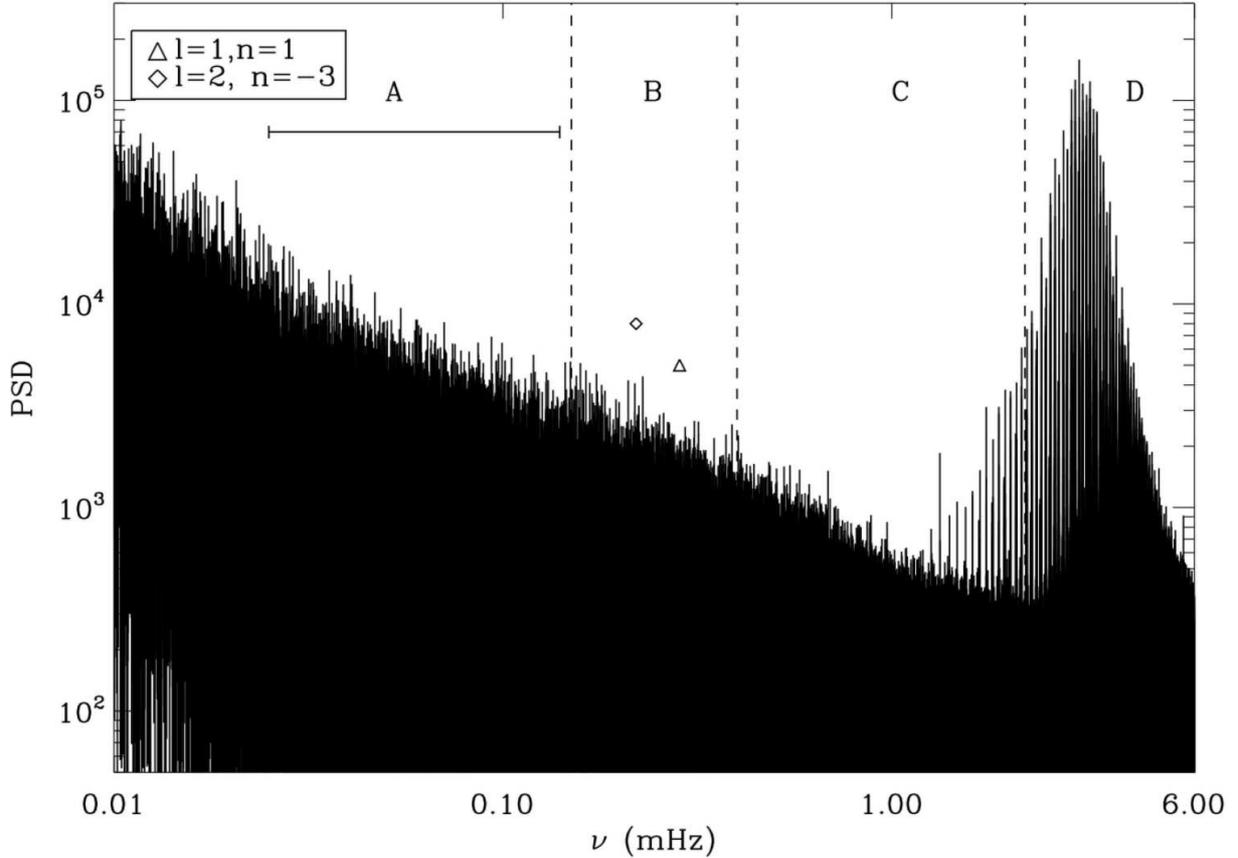}

\end{center}
\caption{\label{Fig1}Power spectrum density of 3660 days of GOLF data. A: region where the asymptotic properties of the g modes have been searched (horizontal segment). B: region where individual g and mixed modes have been looked for. The triangle and the diamond are the two candidates already found. C: region of low-degree low-order p modes obtained with high precision. D: Region of low-degree high order p modes where the blending of the modes produces a reduction in the precision of the retrieved parameters.  }
\end{figure}

For p modes with a given degree $\ell$, the inner turning point $r_t=c_t L/\omega_{\ell,n}$ (where $L=\ell+1/2$, $\omega_{\ell,n}$ is the central frequency of the mode and $c_t$ is a constant \cite{1994A&A...290..845L}) is a decreasing function of frequency. Thus the modes go deeper inside the Sun with increasing frequencies (higher radial order $n$). Unfortunately, the uncertainty (error bars) of the fitted rotational splittings of Sun-as-a-star observations (the most sensitive to low-order p modes) is larger. Indeed, as the lifetime of these modes at high frequency is small, their line width is increased. Therefore, for frequencies above $\sim2.2$ mHz (See region D in Fig.\ref{Fig1}) the error bars of the extracted splittings start to grow up, and above $\sim3.5$ mHz, there is a substantial blending between the visible $m$ components of the p modes making the rotational splitting very difficult to extract. At higher frequencies, even the successive pairs of modes $\ell=$0, 2 and $\ell=$1, 3 are blended together and today it is not possible to obtain values of the rotational splittings with enough accuracy to be useful in the inversion codes. Moreover, there are some sources of biases in the procedure of data fitting that should be treated carefully as, for example, the visibility ratio of the different $m$-components of the modes \cite{ChaApp2006}, or the effect of the magnetic activity cycle on the sectoral components of the modes $\ell \ge 2$ \cite{2003MNRAS.343..343C}.

On the contrary, at low frequency -- below $n$=16, at about 2.2 mHz (See region C in Fig.\ref{Fig1}) -- the lifetime of the modes increases so their line width is very small allowing us to determine their rotational splittings with a very high precision. However, as we have seen previously, these modes have inner turning points at lower depths than the high-frequency modes (above 0.08 and 0.12 $R_\odot$ for the modes $\ell$=1 and 2 respectively). Therefore, even though these modes do not carry any information below $\simeq 0.1 R_\odot$, they would lead to improve our knowledge on the inner rotation rate because they have smaller error bars and they contribute to increase the precision of the inversions \cite{effKor07}.

Today, only those low-degree p modes above $\sim$ 1 mHz have been observed \cite{BerVar2000}\cite{GarReg2001}\cite{GarCor2004}\cite{ChaEls2002} and it is common to restrict the splittings of low-degree modes in the inversions below 2.2 mHz because at higher frequencies the error bars are higher than the fraction of the splitting coming from the layers below 0.2 $R_\odot$ \cite{CouGar2003}.  As a consequence, to better constrain the solar rotation profile using only acoustic modes, we would need to measure reliable modes at low frequency while we need to improve the splitting extraction at high frequencies with smaller error bars. Nowadays, with the data sets of more than 4000 days, it has been shown that we have acquired a higher precision in the extraction of the splittings and, therefore, it is possible to use the low-degree modes up to $\sim$ 3.4 mHz in the inversions (Garc\'\i a, Mathur, Ballot et al. Sol. Phys. submitted).

\section{The quest for gravity modes}

In the early 80's, right after the beginning of the helioseismology, a search dedicated to gravity modes (g modes) started. Several groups looked for both individual modes and the signature of their asymptotic properties  (see for example \cite{1983Natur.306..651D} \cite{1984sss..conf..183F} \cite{1984MmSAI..55...91I} \cite{1988IAUS..123...79P} and the reviews by \cite{HilFro1991} and \cite{1991AdSpR..11...29P}). Unfortunately, none of these candidates could be confirmed as gravity modes by more recent observations. Other attempts have been made to measure g modes outside helioseismology. Indeed, \cite{1995Natur.376..139T} found oscillations in the solar wind that were interpreted as g modes. Unfortunately, later calculations showed that these results could also be compatible with noise \cite{1999ApJ...514..972D}.  Moreover, in the middle of the 90's, two observational opportunities were fully operational in helioseismology: the complete deployment of the ground-base networks (i.e. BISON and GONG) and the launch of SoHO (in particular the GOLF instrument dedicated to look for very low-frequency modes). 
 
With the passage to the new millennium, we attended a new blooming of g-mode research based on the quality and accumulation of data. In 2000, \cite{AppFro2000} looked for individual spikes above 150 $\mu$Hz (see region B in Fig.\ref{Fig1}) in the power spectrum with more than 90\% confidence level not being pure noise.  Although they could not identify any g-mode signature, an upper limit of their amplitudes could be established: at 200 $\mu$Hz, they would be below 10 $mm s^{-1}$ in velocity, and below 0.5 parts per million in intensity. The same year, using the GOLF instrument and different periodogram estimators, a peak was found \cite{2001ESASP.464..473G}\cite{GarReg2001} and it was interpreted as one component of the $\ell=1, n=1$ mixed mode (up to 98\% confidence level). Later, in 2002, Gabriel et al. (2002), using a similar statistical approach, confirmed the existence of this mixed-mode candidate with a longer data set of the same instrument. Besides, two other structures -- already studied with 2 years of GOLF data and reviewed by \cite{GabSTC1999} -- were highlighted to be potentially interesting due to their persistency with time. 

In order to reduce the threshold while maintaining the same confidence level, \cite{STCGar2004} looked
 for multiplets instead of spikes. This research led to several patterns that have been considered
 as g-mode candidates. In particular, a structure around 220 $\mu$Hz retained our attention. 
In fact, \cite{CoxGuz2004} showed theoretically that the g mode with the highest surface amplitude
 would be the $\ell=2, n=-3$ expected at 222.145  $\mu$Hz (from the seismic model computed by 
\cite{CouSTC2003}).

This year, the signature of the dipole modes has been uncovered \cite{2007Sci...316.1591G} analysing the periodogramme of the power spectrum computed between 25 and 140 $\mu$Hz (See region A in Fig.\ref{Fig1}). They found a peak that was attributed to the asymptotic separation for the $\ell$=1 modes. By comparing with solar models and assuming a step rotation profile in the core, they found that the analysis performed favored an average core rotation rate spinning 3 to 5 times faster than the rest of the radiative zone below 0.15$R_\odot$. 

It seems that we will soon have access to these gravity modes, but, how many modes would be needed to better constrain the rotation in the solar core below 0.25 $R_\odot$? The introduction of one g mode in the inversions already improves the obtained rotation rate in the core but the profile is still poorly inferred below 0.16 $R_\odot$ (Mathur, Eff-Darwich, Garc\'\i a \& Turck-Chi\`eze, A\&A submitted). Moreover the inversion including four g modes $\ell$=1 and four $\ell$=2, and depending on the error bars of their splittings, is able to better reproduce the simulated rotation profile in the core while the rate in the deepest layers ($\sim$ 0.1 $R_\odot$) is more accurate.

\section{Conclusions}
A very exciting future is ahead us for the global seismology with the present and future missions that are planned or are already being built. In helioseismology we can mention some new instruments: PICARD, GOLF-NG and SDO and in asteroseismology: Kepler, SONG, PLATO and CoRoT which first data are scheduled to be released at the end of the year. All these instruments, as well as the new analyses techniques that are being developed and the improvements in the theory with better and more accurate solar models would provide a better look inside the Sun and other stars.

\ack
This work has been partially funded by the grant AYA2004-04462 of the Spanish Ministry of Education and Culture. The author wants to thank the European Helio- and Asteroseismology Network (HELAS), a major international collaboration funded by the European Commission's Sixth Framework Programme.

\section*{References}
\bibliography{/Users/rgarcia/Desktop/BIBLIO}

\providecommand{\newblock}{}
\begin{thebibliography}{10}
\expandafter\ifx\csname url\endcsname\relax
  \def\url#1{{\tt #1}}\fi
\expandafter\ifx\csname urlprefix\endcsname\relax\def\urlprefix{URL }\fi
\providecommand{\eprint}[2][]{\url{#2}}

\bibitem{JCD2002}
{Christensen-Dalsgaard} J 2002, {\em Reviews of Modern Physics\/}, {\bf 74},
  1073--1129 (\textit{Preprint} \eprint{astro-ph/0207403})

\bibitem{1991soia.book..401C}
{Christensen-Dalsgaard} J and {Berthomieu} G 1991, {\em {Theory of solar
  oscillations.}\/} (Solar Interior and Atmosphere), 401--478

\bibitem{2003A&A...411..215R}
{Roxburgh} I~W and {Vorontsov} S~V 2003, {\em \aap\/}, {\bf 411}, 215--220

\bibitem{2007ApJ...655..660B}
{Basu} S, {Chaplin} W~J, {Elsworth} Y, {New} R, {Serenelli} A~M and {Verner}
  G~A 2007, {\em \apj\/} {\bf 655}, 660--671 (\textit{Preprint}
  \eprint{arXiv:astro-ph/0610052})

\bibitem{2007arXiv0705.3154C}
{Chaplin} W~J, {Serenelli} A~M, {Basu} S, {Elsworth} Y, {New} R and {Verner}
  G~A 2007, {\em ArXiv e-prints\/}, {\bf 705} (\textit{Preprint}
  \eprint{0705.3154})

\bibitem{1996SoPh..168....1C}
{Chaplin} W~J, {Elsworth} Y, {Howe} R, {Isaak} G~R, {McLeod} C~P, {Miller} B~A,
  {van der Raay} H~B, {Wheeler} S~J and {New} R 1996, {\em \solphys\/}, {\bf 168},
  1--18

\bibitem{HarHil1996}
{Harvey} J~W, {Hill} F, {Hubbard} R, {Kennedy} J~R, {Leibacher} J~W, {Pintar}
  J~A, {Gilman} P~A, {Noyes} R~W, {Title} A~M, {Toomre} J, {Ulrich} R~K,
  {Bhatnagar} A, {Kennewell} J~A, {Marquette} W, {Patr{\'o}n} J, {Sa{\'a}} O
  and {Yasukawa} E 1996, {\em Science\/}, {\bf 272}, 1284

\bibitem{DomFle1995}
{Domingo} V, {Fleck} B and {Poland} A~I 1995, {\em \solphys\/}, {\bf 162}, 1--37

\bibitem{GabGre1995}
{Gabriel} A~H, {Grec} G, {Charra} J, {Robillot} J~M, {Cort\'es} T~R,
  {Turck-Chi\`eze} S, {Bocchia} R, {Boumier} P, {Cantin} M, {C\'espedes} E,
  {Cougrand} B, {Cretolle} J, {Dame} L, {Decaudin} M, {Delache} P, {Denis} N,
  {Duc} R, {Dzitko} H, {Fossat} E, {Fourmond} J~J, {Garc{\'{\i}}a} R~A, {Gough}
  D, {Grivel} C, {Herreros} J~M, {Lagardere} H, {Moalic} J~P, {Pall\'e} P~L,
  {Petrou} N, {Sanchez} M, {Ulrich} R and {van der Raay} H~B 1995, {\em
  \solphys\/}, {\bf 162}, 61--99

\bibitem{1995SoPh..162..101F}
{Frohlich} C, {Romero} J, {Roth} H, {Wehrli} C, {Andersen} B~N, {Appourchaux}
  T, {Domingo} V, {Telljohann} U, {Berthomieu} G, {Delache} P, {Provost} J,
  {Toutain} T, {Crommelynck} D~A, {Chevalier} A, {Fichot} A, {Dappen} W,
  {Gough} D, {Hoeksema} T, {Jimenez} A, {Gomez} M~F, {Herreros} J~M, {Cortes}
  T~R, {Jones} A~R, {Pap} J~M and {Willson} R~C 1995, {\em \solphys\/}, {\bf 162},
  101--128

\bibitem{1995SoPh..162..129S}
{Scherrer} P~H, {Bogart} R~S, {Bush} R~I, {Hoeksema} J~T, {Kosovichev} A~G,
  {Schou} J, {Rosenberg} W, {Springer} L, {Tarbell} T~D, {Title} A, {Wolfson}
  C~J, {Zayer} I and {MDI Engineering Team} 1995, {\em \solphys\/}, {\bf 162},
  129--188

\bibitem{2007MNRAS.379....2B}
{Broomhall} A~M, {Chaplin} W~J, {Elsworth} Y and {Appourchaux} T 2007, {\em
  \mnras\/}, {\bf 379}, 2--10

\bibitem{JimCha2007}
{Jim{\'e}nez-Reyes} S~J, {Chaplin} W~J, {Elsworth} Y, {Garc{\'{\i}}a} R~A,
  {Howe} R, {Socas-Navarro} H and {Toutain} T 2007, {\em \apj\/}, {\bf 654},
  1135--1145

\bibitem{1994A&A...290..845L}
{Lopes} I and {Turck-Chi\`eze} S 1994, {\em \aap\/}, {\bf 290}, 845--860

\bibitem{ChaApp2006}
{Chaplin} W~J, {Appourchaux} T, {Baudin} F, {Boumier} P, {Elsworth} Y,
  {Fletcher} S~T, {Fossat} E, {Garc{\'{\i}}a} R~A, {Isaak} G~R, {Jim\'enez} A,
  {Jim\'enez-Reyes} S~J, {Lazrek} M, {Leibacher} J~W, {Lochard} J, {New} R,
  {Pall\'e} P, {R\'egulo} C, {Salabert} D, {Seghouani} N, {Toutain} T and
  {Wachter} R 200,6 {\em \mnras\/}, {\bf 369}, 985--996 (\textit{Preprint}
  \eprint{astro-ph/0606748})

\bibitem{2003MNRAS.343..343C}
{Chaplin} W~J, {Elsworth} Y, {Isaak} G~R, {Miller} B~A, {New} R, {Thiery} S,
  {Boumier} P and {Gabriel} A~H 2003, {\em \mnras\/}, {\bf 343}, 343--352

\bibitem{effKor07}
{Eff-Darwich} A, {Korzennik} S~G, {Jim\'enez-Reyes} S~J and {Garc\'\i a} R~A,
  2007, {\em \apj\/}

\bibitem{BerVar2000}
{Bertello} L, {Varadi} F, {Ulrich} R~K, {Henney} C~J, {Kosovichev} A~G,
  {Garc{\'{\i}}a} R~A and {Turck-Chi\`eze} S 2000, {\em \apjl\/}, {\bf 537},
  L143--L146

\bibitem{GarReg2001}
{Garc\'\i a} R~A, {R\'egulo} C, {Turck-Chi\`eze} S, {Bertello} L, {Kosovichev}
  A~G, {Brun} A~S, {Couvidat} S, {Henney} C~J, {Lazrek} M, {Ulrich} R~K and
  {Varadi} F 2001, {\em \solphys\/}, {\bf 200}, 361--379

\bibitem{GarCor2004}
{Garc{\'{\i}}a} R~A, {Corbard} T, {Chaplin} W~J, {Couvidat} S, {Eff-Darwich} A,
  {Jim\'enez-Reyes} S~J, {Korzennik} S~G, {Ballot} J, {Boumier} P, {Fossat} E,
  {Henney} C~J, {Howe} R, {Lazrek} M, {Lochard} J, {Pall{\'e}} P~L and
  {Turck-Chi\`eze} S 2004, {\em \solphys\/}, {\bf 220}, 269--285

\bibitem{ChaEls2002}
{Chaplin} W~J, {Elsworth} Y, {Isaak} G~R, {Marchenkov} K~I, {Miller} B~A, {New}
  R, {Pinter} B and {Appourchaux} T 2002, {\em \mnras\/}, {\bf 336}, 979--991

\bibitem{CouGar2003}
{Couvidat} S, {Garc{\'{\i}}a} R~A, {Turck-Chi\`eze} S, {Corbard} T, {Henney}
  C~J and {Jim\'enez-Reyes} S 2003, {\em \apjl\/}, {\bf 597}, L77--L79
  (\textit{Preprint} \eprint{astro-ph/0309806})

\bibitem{1983Natur.306..651D}
{Delache} P and {Scherrer} P~H 1983, {\em \nat\/}, {\bf 306}, 651--653

\bibitem{1984sss..conf..183F}
{Fr{\"o}hlich} C and {Delache} P 1984, {\em Social Studies of Science\/},  183

\bibitem{1984MmSAI..55...91I}
{Isaak} G~R, {van der Raay} H~B, {Palle} P~L, {Cortes} T~R and {Delache} P 1984,
  {\em Memorie della Societa Astronomica Italiana\/}, {\bf 55}, 91--97

\bibitem{1988IAUS..123...79P}
{Pall\'e} P~L and {Roca-Cort\'es} T 1988, {\em Advances in Helio- and
  Asteroseismology\/}, ({\em IAU Symposium\/} vol 123) ed
  {Christensen-Dalsgaard} J and {Frandsen} S, 79

\bibitem{HilFro1991}
{Hill} H, {Fr\"ohlich} C, {Gabriel} M and {Kotov} V~A 1991, {\em {Solar gravity
  modes}\/} (Solar interior and atmosphere (A92-36201 14-92).~Tucson, AZ,
  University of Arizona Press, 1991, p.~562-617.), 562--617

\bibitem{1991AdSpR..11...29P}
{Pall\'e} P~L 1991, {\em Advances in Space Research\/}, {\bf 11}, 29--38

\bibitem{1995Natur.376..139T}
{Thomson} D~J, {Maclennan} C~G and {Lanzerotti} L~J 1995, {\em \nat\/}, {\bf 376},
  139--+

\bibitem{1999ApJ...514..972D}
{Denison} D~G~T and {Walden} A~T 1999, {\em \apj\/}, {\bf 514}, 972--978

\bibitem{AppFro2000}
{Appourchaux} T, {Fr\"ohlich} C, {Andersen} B, {Berthomieu} G, {Chaplin} W~J,
  {Elsworth} Y, {Finsterle} W, {Gough} D~O, {Hoeksema} J~T, {Isaak} G~R,
  {Kosovichev} A~G, {Provost} J, {Scherrer} P~H, {Sekii} T and {Toutain} T 2000,
  {\em \apj\/}, {\bf 538}, 401--414

\bibitem{2001ESASP.464..473G}
{Garc{\'{\i}}a} R~A, {Bertello} L, {Turck-Chi{\`e}ze} S, {Couvidat} S,
  {Gabriel} A~H, {Henney} C~J, {R{\'e}gulo} C, {Robillot} J~M, {Roca
  Cort{\'e}s} T, {Ulrich} R~K and {Varadi} F 2001, {\em SOHO 10/GONG 2000
  Workshop: Helio- and Asteroseismology at the Dawn of the Millennium\/} ({\em
  ESA Special Publication\/} vol 464) ed {Wilson} A and {Pall{\'e}} P~L,
  473--478

\bibitem{GabSTC1999}
{Gabriel} A~H, {Turck-Chi\`eze} S, {Garc{\'{\i}}a} R~A, {Pall\'e} P~L,
  {Boumier} P, {Thiery} S, {Baudin} F, {Grec} G, {Ulrich} R~K, {Bertello} L,
  {Roca Cort\'es} T and {Robillot} J~M 1999, {\em Advances in Space Research\/},
  {\bf 24}, 147--155

\bibitem{STCGar2004}
{Turck-Chi\`eze} S, {Garc{\'{\i}}a} R~A, {Couvidat} S, {Ulrich} R~K, {Bertello}
  L, {Varadi} F, {Kosovichev} A~G, {Gabriel} A~H, {Berthomieu} G, {Brun} A~S,
  {Lopes} I, {Pall\'e} P, {Provost} J, {Robillot} J~M and {Roca Cort\'es} T
  2004, {\em \apj\/}, {\bf 604}, 455--468

\bibitem{CoxGuz2004}
{Cox} A~N and {Guzik} J~A 2004, {\em \apjl\/}, {\bf 613}, L169--L171

\bibitem{CouSTC2003}
{Couvidat} S, {Turck-Chi\`eze} S and {Kosovichev} A~G 2003, {\em \apj\/}, {\bf
  599}, 1434--1448 (\textit{Preprint} \eprint{astro-ph/0203107})

\bibitem{2007Sci...316.1591G}
{Garc{\'{\i}}a} R~A, {Turck-Chi\`eze} S, {Jim\'enez-Reyes} S~J, {Ballot} J,
  {Pall\'e} P~L, {Eff-Darwich} A, {Mathur} S and {Provost} J 2007, {\em
  Science\/}, {\bf 316}, 1591--1593

\end{thebibliography}


\end{document}